\documentclass[preprint2 ]{aastex62}
\submitjournal{ApJ}

\shorttitle{Sample article}
\shortauthors{Lyu et al.}

\begin{document}

\title{XMM-Newton and NICER measurement of the rms spectrum of the millihertz quasi-periodic oscillations in the neutron-star low-mass X-ray binary 4U 1636$-$53 }

\correspondingauthor{}
\email{lvming@xtu.edu.cn}

\author{Ming Lyu}
\affiliation{Department of Physics, Xiangtan University, Xiangtan, Hunan 411105, China}
\affiliation{Key Laboratory of Stars and Interstellar Medium, Xiangtan University, Xiangtan, Hunan 411105, China}

\author{Guobao Zhang}
\affiliation{Yunnan Observatories, Chinese Academy of Sciences (CAS), Kunming 650216, P.R. China}
\affiliation{Key Laboratory for the Structure and Evolution of Celestial Objects, CAS, Kunming 650216, P.R. China}

\author{Mariano M\'endez}
\affiliation{Kapteyn Astronomical Institute, University of Groningen, PO BOX 800, NL-9700 AV Groningen, the Netherlands}

\author{D. Altamirano}
\affiliation{Physics \& Astronomy, University of Southampton, Southampton, Hampshire SO17 1BJ, UK}

\author{G. C. Mancuso}
\affiliation{Instituto Argentino de Radioastronom\'{\i}a (CCT-La Plata, CONICET; CICPBA), C.C. No. 5, 1894 Villa Elisa, Argentina}
\affiliation{Facultad de Ciencias Astron\'omicas y Geof\'{\i}sicas, Universidad Nacional de La Plata, Paseo del Bosque s/n, 1900 La Plata, Argentina}

\author{Fu-Yuan Xiang}
\affiliation{Department of Physics, Xiangtan University, Xiangtan, Hunan 411105, China}
\affiliation{Key Laboratory of Stars and Interstellar Medium, Xiangtan University, Xiangtan, Hunan 411105, China}

\author{Huaping Xiao}
\affiliation{Department of Physics, Xiangtan University, Xiangtan, Hunan 411105, China}
\affiliation{Key Laboratory of Stars and Interstellar Medium, Xiangtan University, Xiangtan, Hunan 411105, China}




\begin{abstract}

We used two XMM-Newton and six Neutron Star Interior Composition Explorer (NICER) observations to investigate the fractional rms amplitude of the millihertz quasi-periodic oscillations (mHz QPOs) in the neutron-star low-mass X-ray binary 4U 1636--53. We studied, for the first time, the fractional rms amplitude of the mHz QPOs vs. energy in 4U 1636--53 down to 0.2 keV. We find that, as the energy increases from $\sim$0.2 keV to $\sim$3 keV, the rms amplitude of the mHz QPOs increases, different from the decreasing trend that has been previously observed above 3 keV. This finding has not yet been predicted by any current theoretical model, however, it provides an important observational feature to speculate whether a newly discovered mHz oscillation originates from the marginally stable nuclear burning process on the neutron star surface.

\end{abstract}



\section{Introduction}

A distinct class of quasi-periodic oscillations (QPOs) was discovered by \citet{revni01} in the neutron-star low-mass X-ray binaries (LMXB) 4U 1636$-$53, 4U 1608$-$52 and Aql X$-$1. The typical frequency of these QPOs is $\sim$5-14 mHz  \citep{revni01,diego08,lyu15,strohmayer18,Mancuso19}, and the QPOs become undetectable once there is onset of a type I X-ray burst \citep{revni01,diego08}. The mHz QPOs are present only when the source luminosities is within a narrow range, $L_{\rm 2-20\:keV} \simeq (5-11) \times 10^{36}$ ergs s$^{-1}$ \citep{revni01,diego08}, and are more significant at low energies ($<$ 5 keV). \citet{revni01} proposed that the mHz QPOs originate from a special mode of nuclear burning on the neutron star surface. This interpretation was consistent with the finding that in 4U 1608--52 the 2--5 keV count rate connected with a 7.5 mHz QPO is anti-correlated with the frequency of the kilohertz (kHz) QPOs \citep{yu02}: the inner disc is ``pushed" outwards by the radiation stresses from the neutron-star surface in each mHz QPO cycle when the luminosity increases, leading to the change of the kHz QPO frequency. In the work of \citet{diego08}, it is found that the mHz QPOs in 4U 1636$-$53 show a systematically decreasing frequency before a type I X-ray burst when the source was in the transitional state. A similar behavior was also reported in the LMXB EXO 0748--676 by \citet{Mancuso19}. These frequency drift indicates that there is a close connection between the QPOs and the nuclear burning on the neutron star surface.
 
Calculations in \citet{heger07} suggest that the mHz QPOs originates from marginally stable nuclear burning of Helium on the neutron-star surface. Their simulation shows an oscillatory mode of burning at a characteristic time scale of $\sim$100 seconds, consistent with the $\sim$ 2-minute period of the mHz QPOs \citep{heger07}. The burning is oscillatory only when the accretion rate close to the Eddington rate, one order of magnitude bigger than the global accretion rate implied from observations. \citet{keek09} found that the turbulent chemical mixing of the fuel, together with a higher heat flux from the crust, is able to generate the mHz QPOs at the observed accretion rate. In their simulation, the frequency drift of the QPOs could be triggered if there is a cooling process of the burning layer. \citet{keek14} further studied the influence of the nuclear reaction rate and the fuel composition in producing the mHz QPOs. They found that, at the observed accretion rate, the mHz QPOs could not be triggered by changing only the composition and the reaction rate.

Millihertz QPOs with different observational properties were reported by \citet{linaries10} in the neutron star transient source IGR J17480$-$2446. These so-called `high-luminosity' mHz QPOs have a frequency of $\sim$4.5 mHz, and the persistent luminosity of the source when the mHz QPOs were present was relatively high, $L_{2-50{\rm keV}}$ $\sim$ 10$^{38}$ erg s$^{-1}$. Interestingly, as the accretion rate increased, type I X-ray bursts in IGR J17480$-$2446 gradually evolved into a mHz QPO, and vice versa \citep{linaries12}.

More recently, \citet{stiele16} found that the oscillations in 4U 1636$-$53 were not consistent with the variations of the temperature of the neutron star surface, whereas \citet{strohmayer18} found that the oscillations in GS 1826$-$238 were due to the blackbody temperature modulation, assuming a constant blackbody normalization in the oscillation cycles. \citet{lyu15} found that, in 4U 1636$-$53, the frequency of the mHz QPOs is not significantly correlated with the temperature of the neutron-star surface, different from theoretical predictions \citep{heger07,keek09}. \citet{lyu16} showed that the 39 type I X-ray bursts associated with mHz QPOs in 4U 1636$-$53 all show positive convexities \citep{Maurer08} and short rising time \citep{Mahmoodifar16}, indicating that the mHz QPOs in this source originate at the equatorial region of the neutron star surface. The finding of \citet{lyu16} also suggests that the local mass accretion rate, from an equatorial accretion disc, could be higher than the global averaged accretion rate, possibly offering a solution to the apparent discrepancy between the model predictions and the observations. \citet{lyu19} investigated the mHz QPOs in 4U 1636--53 using all available Rossi X-ray Timing Explorer (RXTE) observations and found that there was no mHz QPO when the source was in the hard state. Furthermore, \citet{lyu19} found that the absolute RMS amplitude of the mHz QPOs was independent of the parameter S$_{a}$, which is assumed to be an increasing function of the accretion rate \citep{hasinger89,mendez99,zhang11}.

In this paper we studied, for the first time, the amplitude of the mHz QPOs vs. energy in 4U 1636--53 down to 0.2 keV with XMM-Newton and Neutron Star Interior Composition Explorer (NICER) observations. The paper is organized as follows: We describe the observations and the details of the data reduction and analysis in Section 2. In Section 3 we show results derived in this work. In Section 4, we discussed the findings in the frame of marginally stable nuclear burning process.

\section{Observations and data reduction}
In this work we used data taken from XMM-Newton and NICER. The XMM-Newton observations were performed on 2008 February 27 (ObsID: 0500350401, label `X1') and 2009 March 14 (ObsID: 0606070101, label `X2') using the European Photon Imaging Camera, EPIC-PN \citep{xmm01}, in timing mode. Millihertz QPOs have already been reported in these two observations in previous works \citep{lyu15, stiele16}. 

We used the Science Analysis System (SAS) version 16.1.0 for the XMM-Newton data reduction, with the latest calibration files applied. We extracted calibrated events with the tool {\tt epproc}, and applied the command {\tt barycen} to convert the arrival time of photons from the local satellite frame to the barycenter of the solar system. We further applied the SAS task {\tt epiclccorr} to correct the EPIC source time series. We applied the test {\tt epatplot} and found that there was moderate pile up in the X1 and X2. We then selected a 41-column wide region centered at the position of the source, and excluded the central three and one columns for X1 and X2 observation, respectively. We selected only single and double events (pattern $\le$4) to extract 1 second light curves.

\begin{figure}
\center
\includegraphics[height=0.42\textwidth]{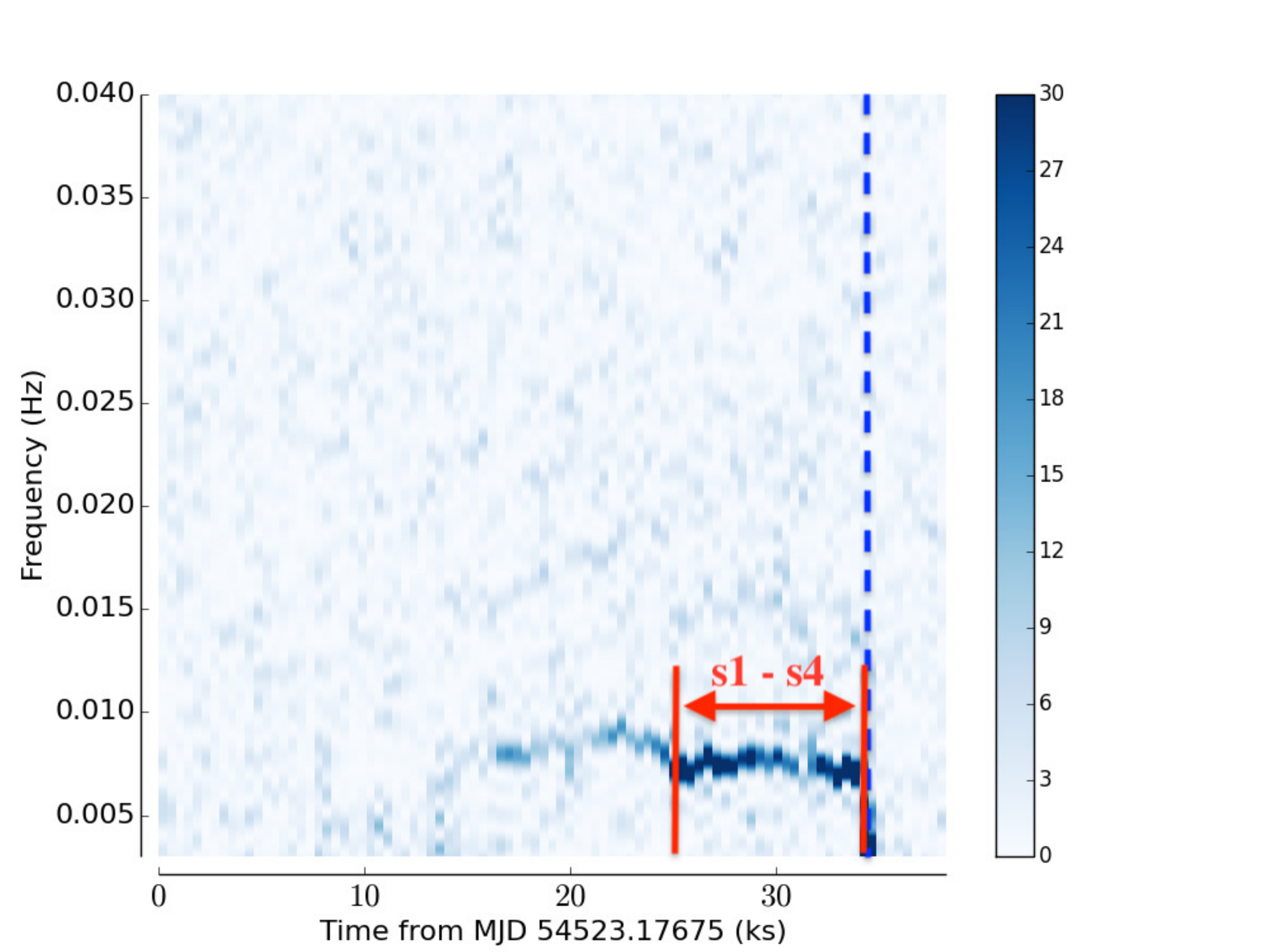}
\includegraphics[height=0.42\textwidth]{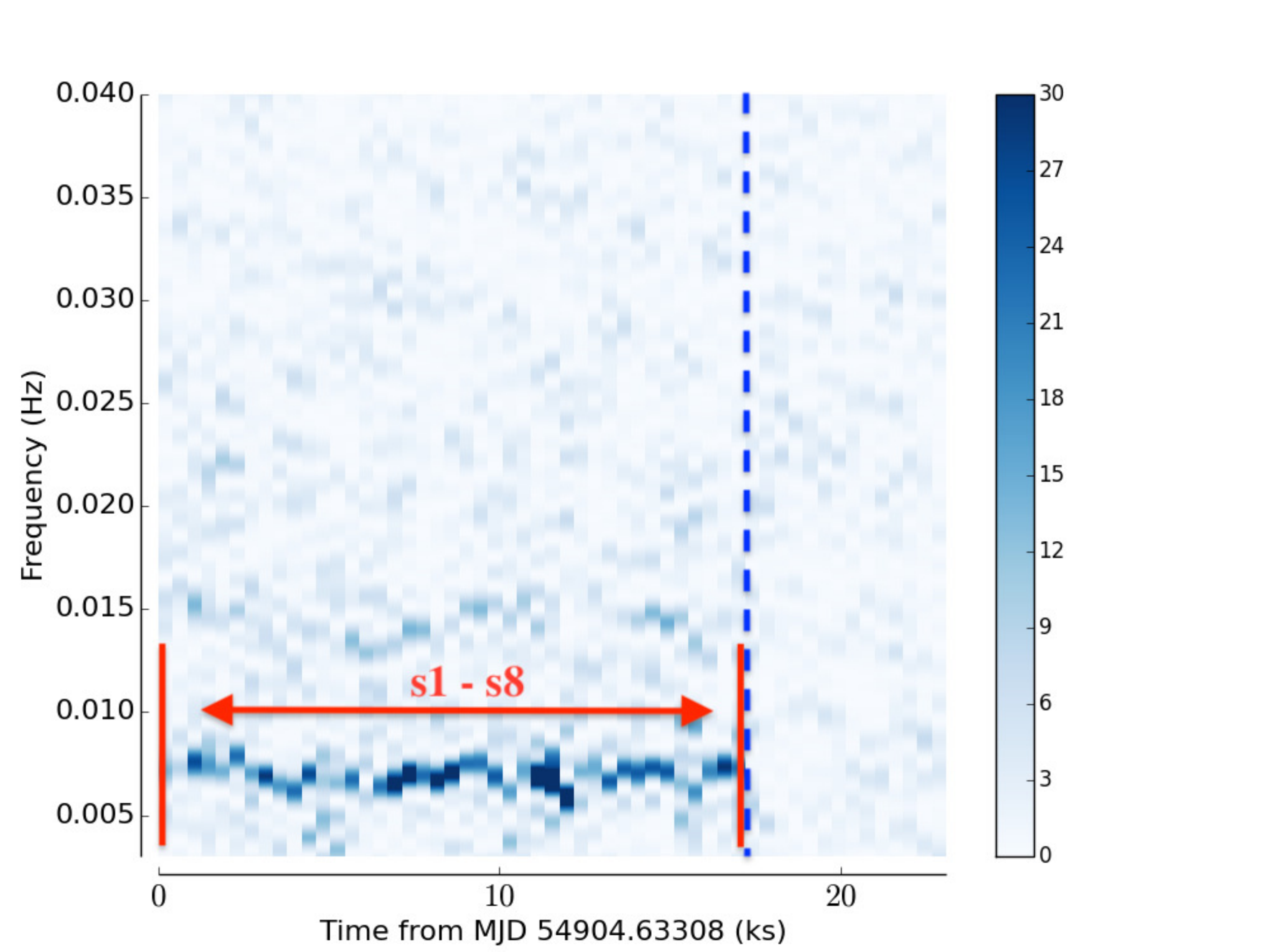}
\caption{Dynamic power spectra of the two XMM-Newton observations (top: 0500350401; bottom: 0606070101) of 4U 1636--53. Each column represents the power spectrum generated from a 840 seconds time interval with the starting time of each interval set to 420 s after the starting time of the previous one. To display the frequency evolution, the frequency is oversampled by a factor of 100 using the Lomb-Scargle periodogram. We fixed the count rate within instrument dropouts and X-ray bursts at the average rate of the whole observation. The colour bars on the right indicate the power at each frequency. We marked the range of the dataset D1 (s1-s4) and D2 (s1-s8) used in this work. The blue dashed line indicates the time of the oneset of a type I X-ray burst in each observation.}
\label{dps}
\end{figure}

We excluded instrument dropouts and X-ray bursts, and produced a light curve from 0.5 keV to 5.3 keV for each XMM-Newton observation. We made dynamic power spectra of these two XMM-Newton observations (Figure \ref{dps}) to locate the time interval during which the mHz QPOs were present. The mHz QPOs are not always present in these two observations and the frequency of the QPO shows clear changes with time. We selected the dataset D1, time interval 24860 s - 33052 s from the start of the observation in X1, and D2, time interval 0 s - 16384 s from the start of the observation in X2, where the variation of the QPO frequency is less than 1 mHz (see Table \ref{t} and Figure \ref{dps} for more details). We then divided the dataset D1 and D2 into several  2048-second segments (s1-s4 in X1; s1-s8 in X2, see Table \ref{t} for more information), and extracted light curves in different energy bands (0.5-1.3 keV; 1.3-2.1 keV; 2.1-2.9 keV; 2.9-3.7 keV; 3.7-4.5 keV; 4.5-5.3 keV) for all the segments.

\begin{figure}
\includegraphics[height=0.4\textwidth,angle=-90]{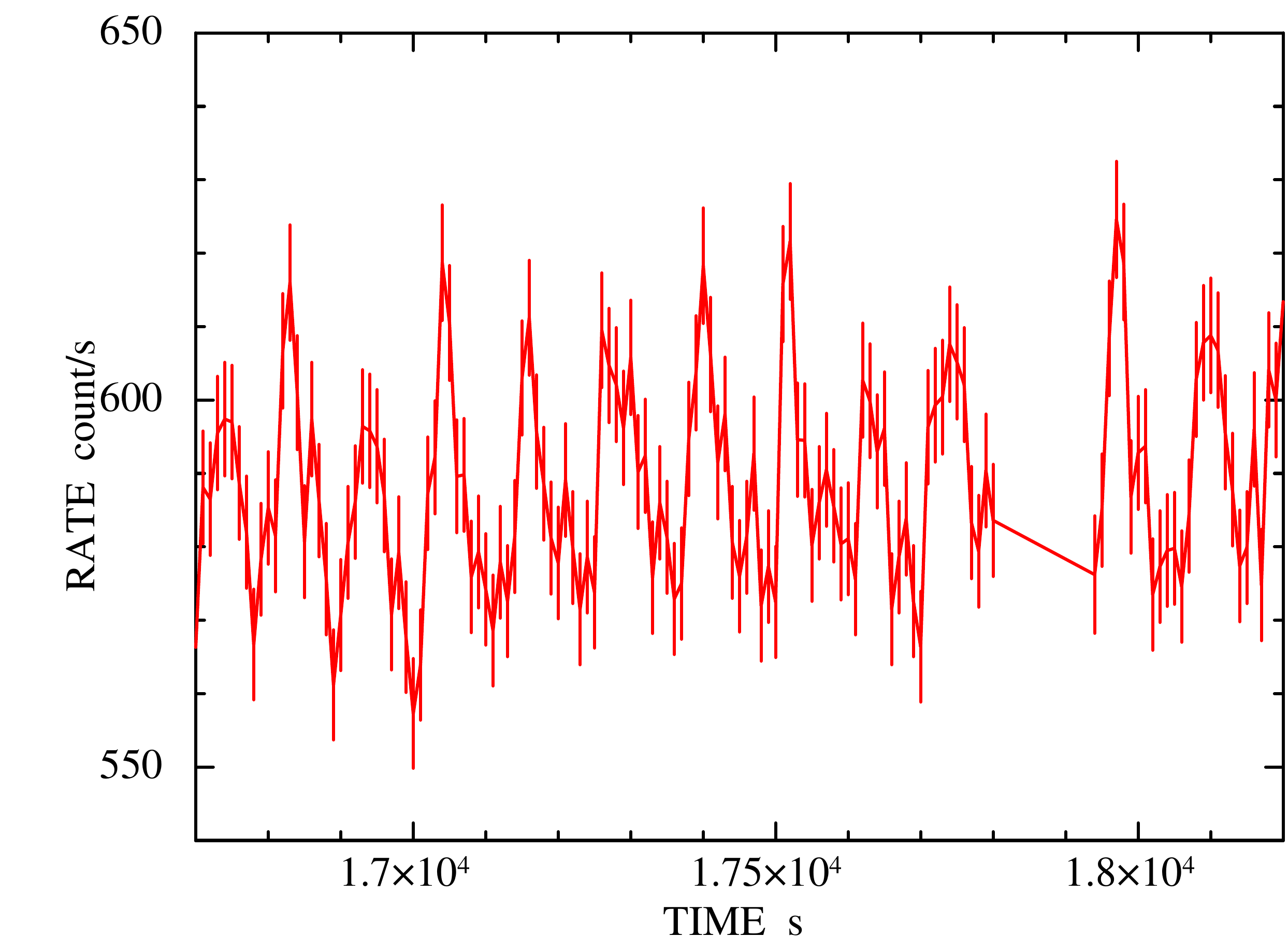}
\includegraphics[height=0.65\textwidth]{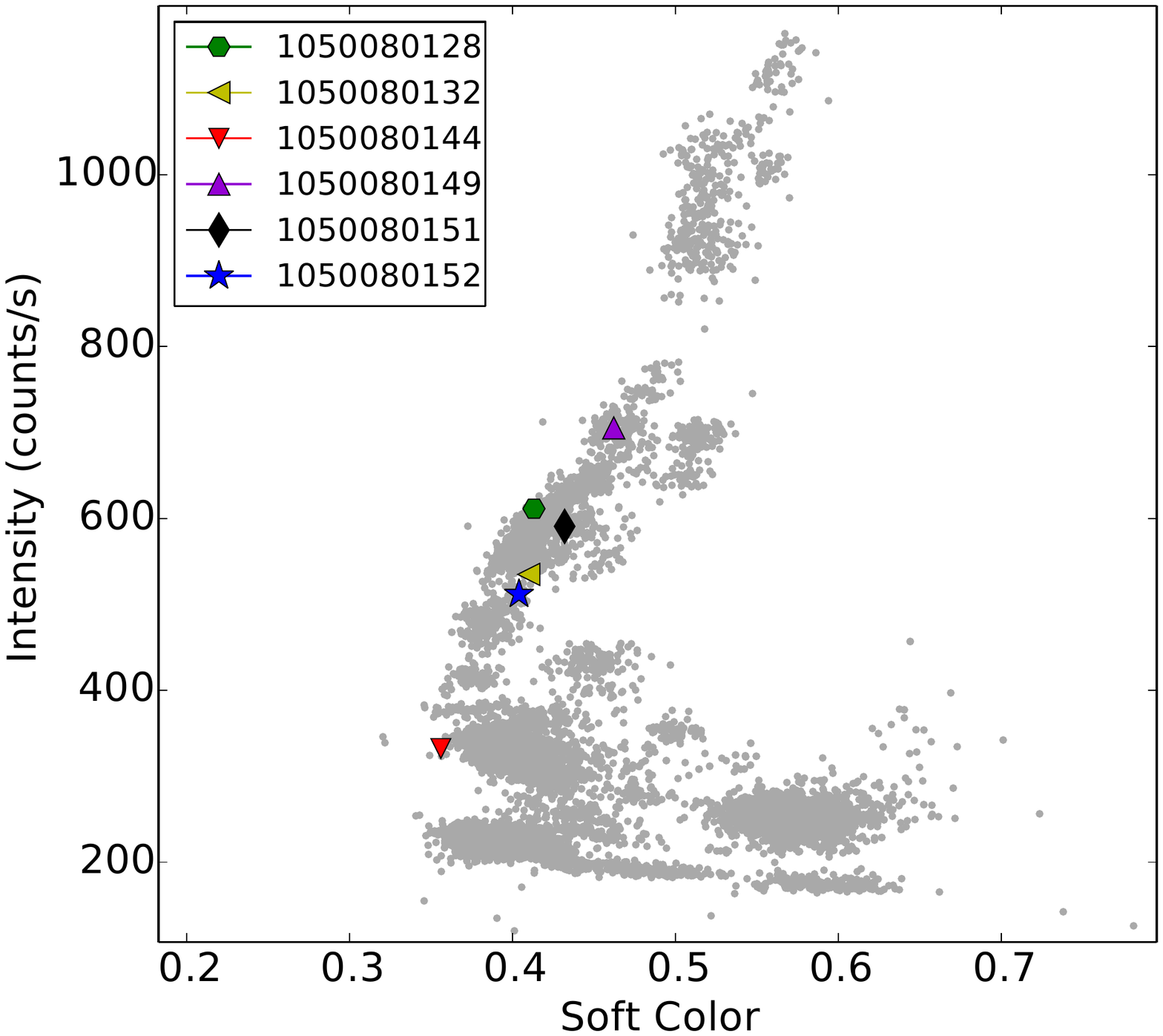}

\caption{Light curve (ObsID:1050080132) and soft-color--intensity diagram (SID) of 4U 1636--53 with NICER. The light curve is in the 0.2-5.0 keV energy, with a time resolution of 10 seconds. Significant quasi-periodic oscillations are present at a timescale of $\sim$110 seconds. The soft color in the SID was computed as the ratio of count rates between the 1.8-3.5 keV and the 0.5-1.8 keV photons using 32 s intervals, while the Intensity was calculated as the count rate in 0.5-6.8 keV.}
\label{lc}
\end{figure}

We analyzed all 124 NICER observations of 4U 1636--53 available in the archive at the time we wrote this paper. The NICER data were processed following standard procedures using the NICER Data Analysis Software NICERDAS 2018-10-07 V005, together with HEASOFT version 6.25. We cleaned the data using standard calibration process with {\tt nicercal} and applied standard screening with {\tt nimaketime} in the full level 2 calibration and screening pipeline {\tt nicerl2}. We then extracted a light curve in the 0.2-5.0 keV range at a 1-s resolution for each NICER observation using {\tt xselect} and searched them for mHz oscillations using Lomb-Scargle periodograms \citep{lomb76,scargle82}. Among all NICER observations, six of them show significant mHz QPOs (see e.g., Fig \ref{lc}). The significance, as estimated from the Lomb-Scargle periodogram, are in all cases above the 3 $\sigma$ confidence level, taking into account the number of trials. For these six observations, we then extracted light curves in different energy bands (0.2-1.0 keV; 1.0-1.8 keV; 1.8-2.6 keV; 2.6-3.4 keV; 3.4-4.2 keV; 4.2-5.0 keV).

As with each XMM-Newton observation, we fitted the 0.5-5.3 keV light curve with mHz QPOs with a model consisting of a sine function plus a constant to get general properties of the QPOs. We then used the best-fit period to fold the light curves in different energy bands using the ftool {\bf efold}. For the NICER observations, we applied the same procedure to fit the 0.2-5.0 keV light curves and folded light curves in different energy bands. 

Besides, we made a soft-color-intensity diagram (SID) to trace the spectral state of the source in these six NICER observations. The soft color was computed as the ratio of count rates in the bands 1.8-3.5 keV and 0.5-1.8 keV using 32 s intervals, while the intensity was calculated as the rate in the 0.5-6.8 keV \citep{Bult18}. 

We then evaluated the influence of the background in the calculation of fractional rms amplitude. For the XMM-Newton data in timing mode, the whole CCD was contaminated by the source photons due to the wide point spread function (PSF) of the telescope \citep{Ng10,hiemstra11,sanna13}. To extract the background, we then selected the observation of 4U 1608--52 (ObsID 0074140201) in the same observational mode when the source was close to quiescence \citep{lyu19a}. The background is extracted from a region RAWX in [17:57] without including the time interval with flaring particle. The derived ratio of the background count rate to the total rate in each energy band is very small, $\sim$ 0.2\%. On the other hand, the NICER background was estimated to be $\sim$ 0.5 cts/s/keV at $\sim$0.6 keV and $\sim$ 0.1 cts/s/keV above $\sim$1.4 keV \citep{keek18}. Therefore, in the following analysis we do not take the background into account in the calculations since it is smaller than the errors of the fractional rms amplitude.

Finally, we fitted each folded light curve with a function consisting of a constant term plus a sine function with the period fixed at 1, and calculated the fractional rms amplitude of the mHz QPO in different bands, $rms=A /[\sqrt{2}*C]$, where A is the amplitude of the sine function and C is the value of the constant component. For the two XMM-Newton observations, the final rms amplitude is then derived as the average of the rms amplitudes in all time segments. For the NICER observations, we calculated the rms amplitude directly in each observation since the time segments with mHz QPO are very short, less than $\sim$ 1500 s.

\section{Results}
In Table \ref{t}, we show the average frequency and the fractional rms amplitude of the mHz QPOs in the 0.5-5.3 keV and 0.2-5.0 keV energy bands for, respectively, the segments in the XMM-Newton and NICER observations. The average frequency of the mHz QPOs in the XMM-Newton observations was in the range 7.2-7.7 mHz and 6.6-7.5 mHz for the observation X1 and X2 in the 0.5-5.3 keV energy band, respectively. The fractional rms amplitude of the QPOs ranges from 0.85\% to 1.32\% in X1, and from 0.69\% to 1.05\% in X2. The mHz QPOs in the NICER observations cover a relative wide frequency range, 5.2-11.1 mHz, with the rms amplitude in the 0.2-5.0 keV band being between 1.10\% and 3.79\%. In Figure \ref{lc}, we show the distribution of the six NICER observations with mHz QPOs in the soft-color--intensity diagram. Similar to the colour diagram in \citet{zhang11} and \citet{Bult18}, the Figure shows that the source went from the soft spectral state to the transitional spectral state as it moved from the top to the left bottom in the diagram, and then the source moved to the hard spectral state when it went to the right bottom. The intensity and the soft color for these six NICER observations are in the range 300-800 cts/s and 0.35-0.5, respectively, when the source was in the intermediate/transitional spectral state \citep[see, e.g.][]{Bult18}.

\begin{figure}
\center
\includegraphics[height=0.45\textwidth]{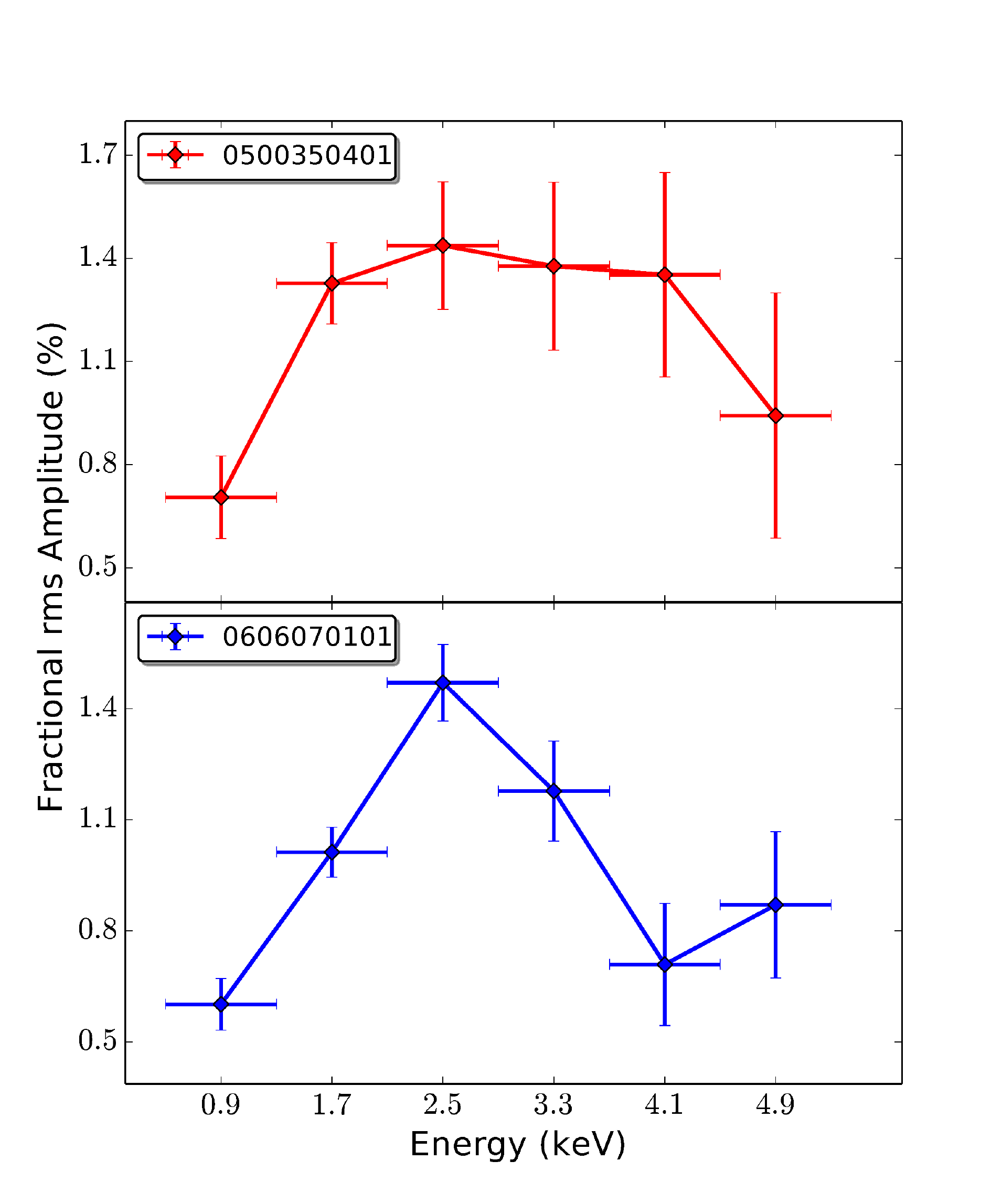}
\caption{Fractional rms amplitude of the mHz QPOs in 4U 1636--53 as a function of energy for the two XMM-Newton observations (X1 to X2 from top to bottom). The energy in the plot represents the central energy of each band, with the error bar indicating the energy range of the band.}
\label{result_xmm}
\end{figure}

\begin{figure}
\center
\includegraphics[height=0.8\textwidth]{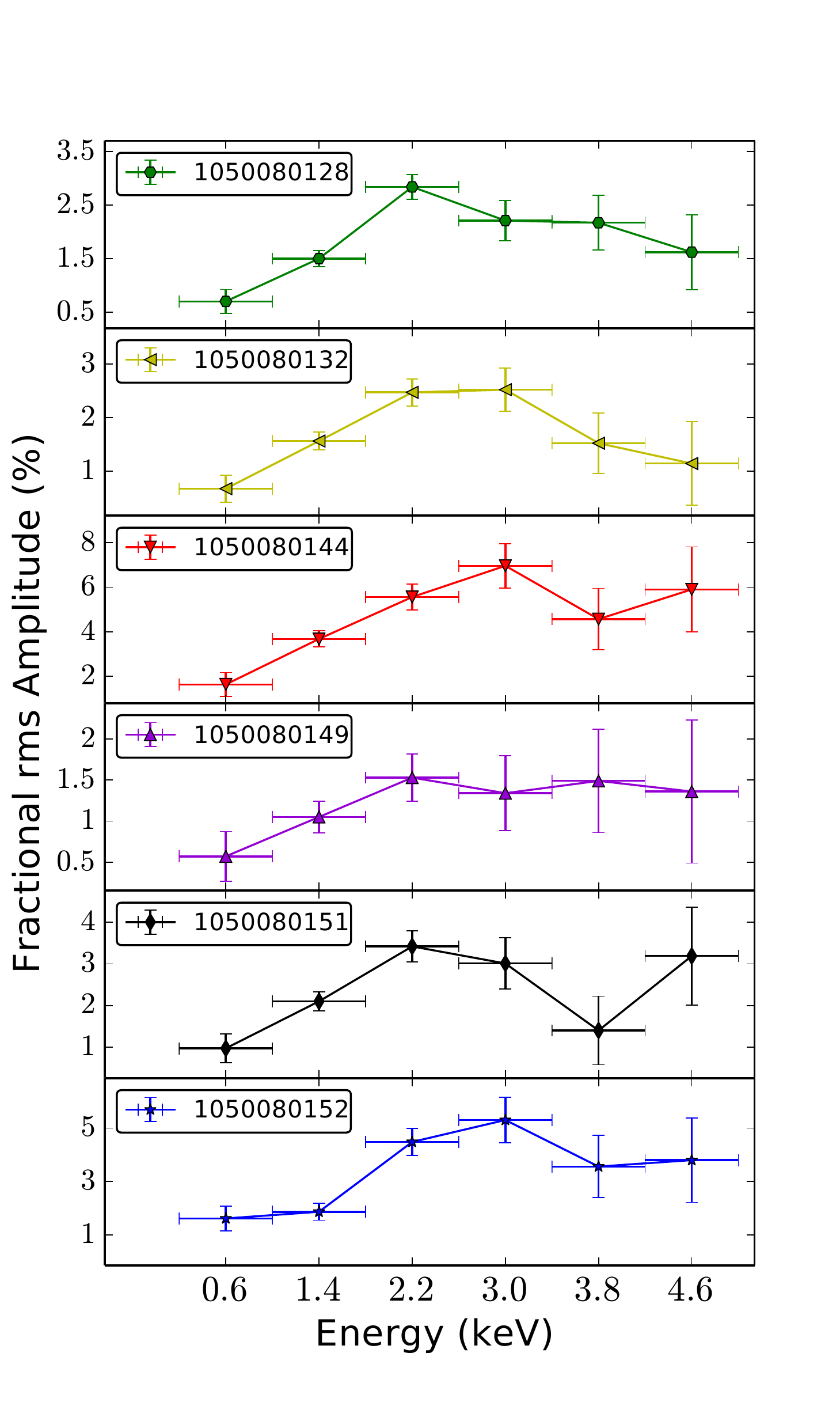}
\caption{Fractional rms amplitude of the mHz QPOs in 4U 1636--53 as a function of energy for the NICER observations. The energy in the plot represents the central energy of each band, with the error bar indicating the energy range of the band.}
\label{result_nicer}
\end{figure}

In Figure \ref{result_xmm}, we show the fractional rms amplitude of the mHz QPOs vs. energy in the two XMM-Newton observations. The rms amplitude first increases from $\sim$0.6\% at 0.9 keV to $\sim$1.4\% at 2.5 keV, and then decreases as the energy further increases. The mHz QPOs in the NICER observations follow a similar trend, although errors of the fractional rms amplitude at energies $>$ 3 keV in some observations are relatively large. As shown in Figure \ref{result_nicer}, the rms amplitude of the QPO in each NICER observation generally increases from 0.2 keV to 2.2-3.0 keV, and then decreases or remains more or less constant as the energy further increases up to 5.0 keV.

\begin{table*}
\small
\centering
\caption{XMM-Newton and NICER observations of the mHz QPOs in 4U 1636--53. All errors in the table are at the 68\% confidence level. The rms amplitude is measured in the 0.5-5.3 keV and the 0.2-5.0 keV band for the two XMM-Newton (X1,X2) and the NICER observations, respectively. In the two XMM-Newton observations, the errors of the rms amplitude in different segments in the same observation are the same after being rounded off at two decimals.}
\begin{tabular}{|c|c|c|c|c|}
\hline
Observation ID   &   Segment     & Time range (s)  & Average frequency (mHz)  &  rms amplitude (\%)  \\
\hline
0500350401 (X1)   & s1   & 24860-26908 &     7.22$\pm$0.03 &  1.32$\pm$0.14   \\
                               & s2   & 26908-28956 &     7.49$\pm$0.04 &  0.98$\pm$0.14   \\
                               & s3   & 28956-31004 &     7.63$\pm$0.04 &  0.99$\pm$0.14   \\
                               & s4   & 31004-33052 &     7.27$\pm$0.04 &  0.85$\pm$0.14   \\

\hline
0606070101 (X2)  & s1  & 0-2048            &  7.35$\pm$0.03 & 0.85$\pm$0.11    \\
                              & s2  & 2048-4096      &  7.13$\pm$0.04 & 0.69$\pm$0.11   \\
                              & s3  & 4096-6144      &  6.66$\pm$0.03 & 0.88$\pm$0.11    \\
                              & s4  & 6144-8192      &  6.77$\pm$0.03 & 1.00$\pm$0.11    \\
                              & s5  & 8192-10240    &  7.43$\pm$0.03 & 0.92$\pm$0.11    \\
                              & s6  & 10240-12288  &  6.80$\pm$0.03 & 1.05$\pm$0.11    \\
                              & s7  & 12288-14336  &  7.08$\pm$0.03 & 0.94$\pm$0.11    \\           
                              & s8  & 14336-16384  &  7.43$\pm$0.04 & 0.78$\pm$0.11    \\           

\hline

\hline         
1050080128 & - &  11300-12950  &     9.18$\pm$0.02    &  1.68$\pm$0.10  \\           
1050080132 & - &  16700-18200  &     8.71$\pm$0.03    &  1.61$\pm$0.11  \\     
1050080144 & - &        0-480  &     9.06$\pm$0.07    &  3.79$\pm$0.25  \\           
1050080149 & - &  48490-49340  &     7.80$\pm$0.08    &  1.10$\pm$0.13  \\   
1050080151 & - &  38760-39570  &     5.26$\pm$0.05    &  2.13$\pm$0.16  \\           
1050080152 & - &  22460-22930  &    11.16$\pm$0.10    &  2.60$\pm$0.22  \\

\hline
\end{tabular}
\medskip  
\label{t}
\end{table*}

\begin{figure}
\center
\includegraphics[height=0.45\textwidth]{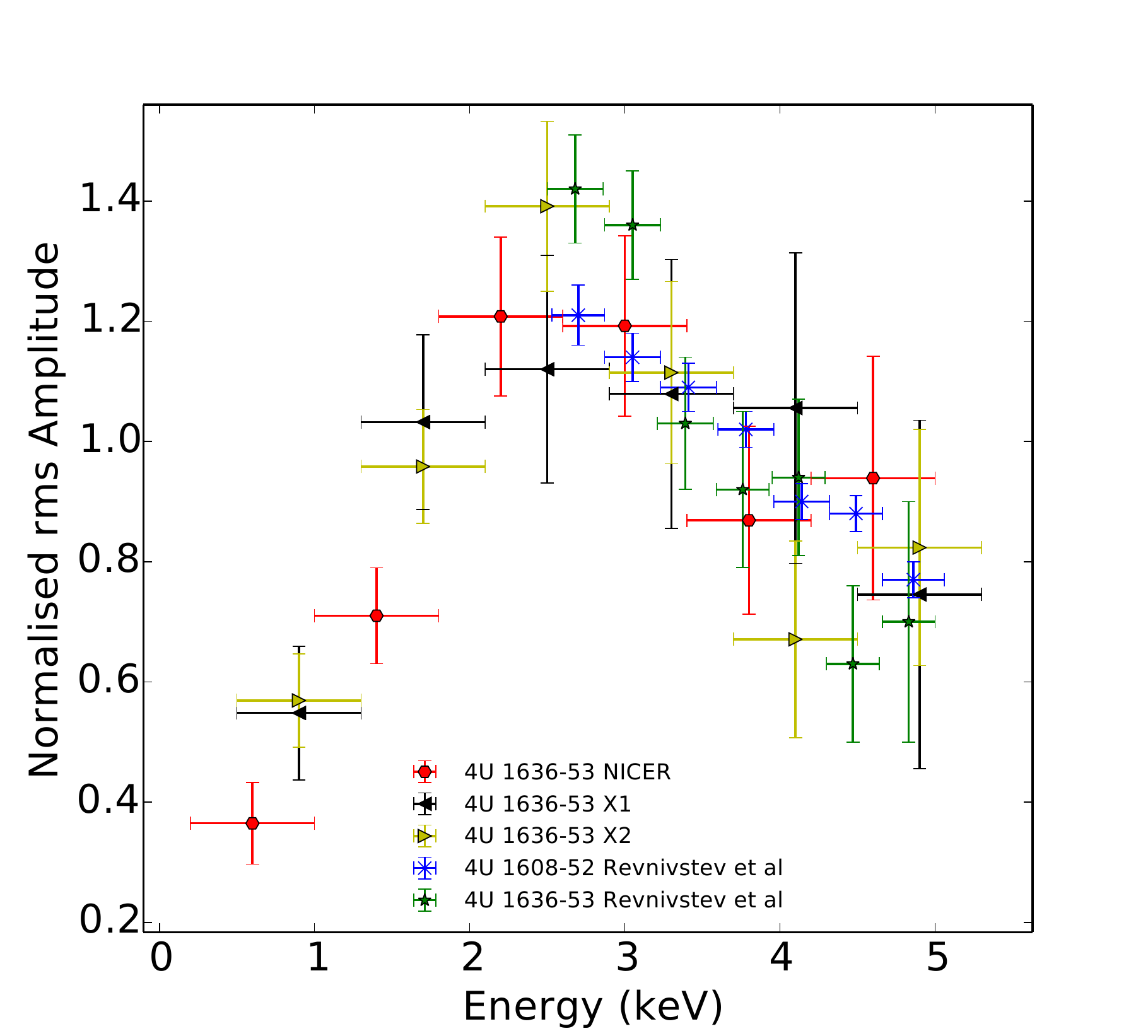}
\caption{Normalised rms amplitude of the mHz QPOs as a function of energy for the XMM-Newton and the NICER observations in this work plus the RXTE observations in \citet{revni01}. We rescaled all the rms spectra of the mHz QPOs in the NICER observations by their average and calculated the mean of the rescaled rms at each energy. We then normalised all the rms spectra in this work and in \citet{revni01} by their averages in the 2.5-5 keV range to bring them to the same scale for comparison. The energy in the plot represents the central energy of each band, with the error bar indicating the energy range of the band.}
\label{result_average}
\end{figure}

\section{Discussion}
We studied the fractional rms amplitude of the mHz QPOs vs. energy in 4U 1636--53 with XMM-Newton and NICER. For the first time, we found that the fractional rms amplitude of the mHz QPOs in 4U 1636--53 increases from 0.2 keV to 2.5-3.0 keV, different from the trend observed above 3 keV with RXTE \citep{revni01,diego08}.

With NICER observation, \citet{strohmayer18} found that in GS 1826--238 there is a clear increasing trend of the rms amplitude of the mHz QPOs from 1 to 3 keV. Our results on the XMM-Newton and the NICER observations of 4U 1636--53 show a clear increasing trend up to $\sim$ 3 keV, indicating that the mHz QPOs in 4U 1636--53 have the same trend as in GS 1826--238 below 3 keV. Considering that the mHz QPOs in both 4U 1636--53 and GS 1826--238 show an increasing trend as energy increases up to 3 keV, it is likely that this is an intrinsic feature of this type of QPOs. 

The rms amplitude derived in the XMM-Newton observations decreases as the energy further increases above 3 keV, consistent with the previous results with RXTE observations. Using RXTE data, \citet{revni01} found that in 4U 1608--52 and 4U 1636--53 the fractional rms amplitude decreases with energy from 2.5 keV to 5 keV. In the NICER observations the rms amplitude above 3 keV is consistent with either decreasing or remaining more or less constant. We further divided the rms spectrum of the mHz QPO in each NICER observation by its average for comparison. We found that the rescaled rms spectra in all NICER observations were consistent with being the same within errors. We then calculated the average of these rescaled rms at each energy. Finally, we rescaled the averaged NICER rms spectrum, the XMM-Newton rms spectra and the rms spectra in \citet{revni01} by their averages in the 2.5-5 keV range to bring them to the same scale for comparison. As shown in Figure \ref{result_average}, the normalised rms spectra above $\sim$2.5 keV in this work are consistent with the ones in \citet{revni01}. These normalised rms spectra further confirms that the fractional rms amplitude of the mHz QPOs first increases and then decreases as the energy increases, with the turnover point around 2.5-3.0 keV. 

Thanks to the low-energy coverage provided by NICER and XMM-Newton, we were able to show, for the first time, that there is a change in the rms amplitude vs. energy relation of the mHz QPOs in 4U 1636--53 around 2.5-3.0 keV. The mechanism responsible for this change is still an open question. Existing models connect the rms amplitude variation to the change of the accretion rate and the crust luminosity, however, to our knowledge neither of them could potentially explain the change of the rms spectra found in this work. In the model of \citet{heger07}, as the accretion rate varies, the transition between stable and unstable burning naturally leads to significant changes in the amplitude of the mHz QPOs. Notwithstanding, the work of \citet{lyu19} indicates that in 4U 1636--53 there is no significant correlation between the absolute RMS amplitude and the parameter S$_{a}$, which is assumed to be an increasing function of the accretion rate \citep{hasinger89,mendez99,zhang11}.

\citet{keek09} found that in their model the simulated mHz QPOs exhibit an increasing amplitude when the heat flux from the neutron-star crust decreases, and that the amplitude of oscillatory burning becomes much larger when considering the turbulent chemical mixing of the fuel. Apparently, the energy dependence of the rms amplitude of the mHz QPOs could not be interpreted in this scenario. More work is still needed to understand the mechanism behind this relation. Interestingly, type I X-ray bursts in this source have a colour temperature around 2 keV \citep{zhang11}, which is close to the turnover point of the derived rms amplitude vs. energy relation derived in this work. Both the mHz QPOs and type I X-ray bursts originate from nuclear burning on the neutron-star surface\citep[e.g.,][]{Paczynski83,cumming04,heger07,keek09}, and the model of the mHz QPOs also predicts the evolution between the mHz QPOs and bursts seen in observations (see Figure 5 in \citet{heger07} and Figure 9 in \citet{keek09} for more details). It is possible that both the colour temperature of the bursts and the turnover point of the rms amplitude vs. energy relation are connected to the same physical factor on the neutron-star surface.

\citet{ferrigno17} reported a strong QPO at $\sim$8 mHz in the accreting millisecond X-ray pulsar IGR J00291+5934 in an XMM-Newton observation. The frequency of this 8 mHz QPO did not drift, and the QPO was present throughout the entire observation. \citet{ferrigno17} found that the rms amplitude of the QPO was around 29\% at $\sim$0.7 keV and decreases dramatically as the energy increases, reaching $\sim$ 7\% at around 6-10 keV. The physical origin of this 8 mHz QPO is still uncertain. The possibility that this 8 mHz QPO is connected to marginally stable nuclear burning on the neutron star surface can neither be excluded nor more solidly confirmed. The reason is that there is only one type I X-ray burst ever detected in this source, which occurred about three days before this XMM-Newton observation \citep{df17}. The rms amplitude vs. energy relation derived in this work, and the work of \citet{strohmayer18} may help us decide whether this 8 mHz QPO originates from nuclear burning. The rms amplitude of the mHz QPOs in both 4U 1636--53 and GS 1826--238 increases with energy below 3 keV. Considering that, compared to the mHz QPOs in these two sources, the  8 mHz QPO in IGR J00291+5934 shows a decreasing rms amplitude as energy increases below 3 keV, together with the fact that the rms amplitude in IGR J00291+5934 is at least an order of magnitude larger than the ones in other mHz QPOs, suggests that the 8 mHz QPO in IGR J00291+5934 does not originate from marginally stable nuclear burning on the neutron star surface.

\acknowledgments
This work is based on observations obtained with XMM-Newton, an ESA science mission with instruments and contributions directly funded by ESA Member States and NASA. This research has made use of data obtained from the High Energy Astrophysics Science Archive Research Center (HEASARC), provided by NASA's Goddard Space Flight Center. This research made use of NASA's Astrophysics Data System. Lyu is supported by National Natural Science Foundation of China (grant No.11803025) and the Hunan Provincial Natural Science Foundation (grant No. 2018JJ3483). G.B. acknowledges funding support from the National Natural Science Foundation of China (NSFC) under grant numbers U1838116 and  the CAS Pioneer Hundred Talent Program Y7CZ181002. DA acknowledges support from the Royal Society. GCM and DA acknowledge support from the Royal Society International Exchanges ``The first step for High-Energy Astrophysics relations between Argentina and UK". GCM was partially supported by PIP 0102 (CONICET) and received financial support from PICT-2017-2865 (ANPCyT). F.Y.X. is supported by the Joint Research Funds in Astronomy (U1531108 and U1731106). H.P.X. is supported by National Natural Science Foundation of China (grant No.11473023).

\software{SAS 16.1.0, HEASOFT 6.25, NICERDAS 2018-10-07, Python 3.6.5}
\facility{XMM-Newton, NICER} 

\bibliographystyle{aasjournal}
\bibliography{paper}

\end{document}